\documentclass[10pt,aps,prd,twocolumn,superscriptaddress,preprintnumbers,amsmath,amssymb,nofootinbib,floatfix]{revtex4-2}

\usepackage{graphicx}
\usepackage{xcolor}
\usepackage{amsmath}
\usepackage{amssymb}
\usepackage{bm}
\usepackage{booktabs}
\usepackage{multirow}
\usepackage{hyperref}
\hypersetup{colorlinks=true,citecolor=blue,linkcolor=blue,urlcolor=blue}

\newcommand{\Msun}{M_\odot}
\newcommand{\dd}{\mathrm{d}}
\newcommand{\GR}{\mathrm{GR}}
\newcommand{\br}{\mathrm{br}}
\newcommand{\detf}{\mathrm{det}}
\newcommand{\src}{\mathrm{src}}

\newcommand{\db}{\delta_{\br}}
\newcommand{\Dbr}{\mathcal{D}_{\br}}

\begin{document}

\title{Polarization Birefringence and Waveform Systematics in GW231123}

\author{Tonghua Liu}
\email{liutongh@yangtzeu.edu.cn}
\affiliation{School of Physics and Optoelectronic, Yangtze University, Jingzhou 434023, China;}
\author{Chenggang Shao}
\email{cgshao@hust.edu.cn}
\affiliation{School of Physics and Optoelectronic, Yangtze University, Jingzhou 434023, China;}

\author{Kai Liao}
\affiliation{School of Physics and Technology, Wuhan University, Wuhan 430072, China;}

\date{\today}

\begin{abstract}
GW231123 is a short, massive binary-black-hole event whose source properties show strong waveform dependence. We use this event to test gravitational-wave polarization birefringence, modeled as a frequency-dependent rotation of the tensor-polarization basis. Instead of sampling a distance-normalized coefficient directly, we sample the band-differential rotation \(\db=\Delta(448\,\mathrm{Hz})-\Delta(20\,\mathrm{Hz})\) with prior \([-\pi,\pi]\), and report the derived coefficient \(\beta_{\br}^{\rm derived}\) for comparison with standard propagation parametrizations. We analyze three waveform families: \textsc{IMRPhenomXPHM} (XPHM), \textsc{IMRPhenomXO4a} (XO4a), and \textsc{NRSur7dq4}. The derived posteriors are consistent with the GR value, giving \(90\%\) upper limits \(|\beta_{\br}^{\rm derived}|_{90}=0.378,\,0.097,\,0.273\) for XPHM, XO4a, and NRSur7dq4, respectively. The directly sampled \(\db\) posterior remains broad, with \(|\db|_{90}\simeq2.8\,\mathrm{rad}\), so the accumulated rotation across the analysis band is weakly constrained. The Bayes factors are waveform dependent: \(\ln\mathcal{B}_{\br/\GR}=-1.26\pm0.30\), \(+3.64\pm0.28\), and \(-0.86\pm0.29\), respectively. We therefore find no waveform-robust evidence for parity-violating propagation. The positive XO4a result is better interpreted as a waveform-dependent birefringence-like response associated with the mass-ratio--distance--spin degeneracy of this short high-mass event.
\end{abstract}

\maketitle

\section{Introduction}
\label{sec:intro}

The first direct detections of gravitational waves from compact-binary coalescences opened a new way to test general relativity (GR) in both the strong-field generation of radiation and the propagation of gravitational waves over cosmological distances \cite{2016PhRvL.116f1102A,2019PhRvD.100j4036A,2021PhRvD.103l2002A,2025PhRvD.112h4080A}.  Propagation tests are sensitive to effects that accumulate between the source and the detector, including modified dispersion, anomalous damping, non-GR polarizations, amplitude birefringence, and phase birefringence \cite{2014LRR....17....4W,2012PhRvD..85b4041M,2025PhRvD.112h4080A}.  In parity-violating gravity the two circular tensor polarizations can propagate differently.  In the phase-birefringence case considered here, the accumulated relative phase rotates the usual linear polarizations and changes the detector response.

Phase and amplitude birefringence arise in Chern--Simons-like extensions of GR and in more general parity-violating propagation frameworks \cite{2009PhR...480....1A,2018PhRvD..98l4018N}.  Previous gravitational-wave birefringence searches have mostly used catalogs of compact binaries to constrain a common propagation parameter \cite{2020PTEP.2020i3E01Y,2022PhRvD.106h4005W}, or have used multiple events to constrain helicity-dependent amplitude birefringence in Chern--Simons gravity \cite{2022PhRvD.106d4067O}.  Those analyses answer a population question: whether the observed catalog supports a common deviation from GR.  The present paper asks a different question.  We use one difficult event, GW231123, to test whether a birefringence-like parameter remains stable when the waveform family is changed.

GW231123 is well suited for this purpose.  The LVK analysis reports a confident two-detector detection with network matched-filter signal-to-noise ratio about 20.7, source-frame component masses of order \(10^2\Msun\), large spin magnitudes, and a remnant in the intermediate-mass black-hole range \cite{2025ApJ...993L..25A}.  The signal is short: only about five visible cycles are present, and most of the power is concentrated between roughly \(30\) and \(80\,\mathrm{Hz}\).  The official source analysis uses an 8 s likelihood segment and the frequency interval \(20\)--\(448\,\mathrm{Hz}\).  Hanford uses a BayesWave glitch-subtracted strain frame, while Livingston uses the original strain data because the identified low-frequency glitch does not significantly affect the source analysis \cite{2025ApJ...993L..25A,2015CQGra..32m5012C}.

The event is astrophysically important because its inferred masses lie in or above the pair-instability mass-gap region, depending on waveform model and mass-gap definition \cite{2025ApJ...993L..25A}.  This connects it to earlier high-mass events such as GW190521, where mass-gap and intermediate-mass-black-hole interpretations also played a central role \cite{2020PhRvL.125j1102A,2020ApJ...900L..13A}.  It is also technically difficult.  The LVK analysis reports substantial waveform dependence among precessing higher-mode models, including \textsc{NRSur7dq4}, \textsc{SEOBNRv5PHM}, \textsc{IMRPhenomTPHM} (TPHM), \textsc{IMRPhenomXPHM} (XPHM), and \textsc{IMRPhenomXO4a} (XO4a) \cite{2025ApJ...993L..25A,2023PhRvD.108l4037R,2022PhRvD.105h4040E,2021PhRvD.103j4056P,2024PhRvD.109f3012T,2019PhRvR...1c3015V}.  This matters for beyond-GR tests.  A propagation parameter can improve a likelihood either because the data contain a real propagation effect or because the parameter absorbs missing waveform content.  The modified-dispersion behavior reported in the LVK analysis already illustrates this danger: apparent deviations can depend on the waveform family and need not imply modified propagation \cite{2025ApJ...993L..25A}.

GW231123 is therefore complementary to very loud and comparatively clean events such as GW250114, whose network matched-filter signal-to-noise ratio is about 80 and whose source is near equal mass with small spins \cite{2025PhRvL.135k1403A}.  A clean, high-SNR event is the natural target for a sharp null test.  GW231123 has a different value: it tests a failure mode.  It asks whether a polarization-propagation parameter can look significant in one waveform model and disappear in another.  This is exactly the situation a future propagation test must understand before claiming parity violation.

We implement phase birefringence as a frequency-dependent rotation of the linear tensor-polarization basis.  Our production analyses use XPHM, XO4a, and NRSur7dq4.  The key result is direct: the posteriors are consistent with GR, and the evidence is not stable across waveform families.  XO4a gives positive support for the birefringent extension, but XPHM and NRSur7dq4 do not.  We therefore interpret the XO4a result as a waveform-dependent birefringence-like response, not as evidence for parity-violating propagation.  Recent lensing, microlensing, and overlapping-signal studies of GW231123 reinforce the same lesson: short high-mass signals can produce frequency-dependent structure that is difficult to attribute uniquely with only two detectors \cite{2025arXiv251216916C,2025arXiv250908298L,2025arXiv251217550H,2025arXiv251217631G,2025arXiv251219077C,2025arXiv251219118S}.

\section{Birefringent propagation model}
\label{sec:model}

A GR waveform model provides two tensor polarizations, \(h_+(f)\) and \(h_\times(f)\).  In parity-violating gravity, the two circular tensor polarizations can acquire different propagation phases.  This is the gravitational-wave analogue of birefringence: right- and left-handed tensor helicities do not propagate in exactly the same way.  Such effects arise in Chern--Simons modified gravity and in more general parity-violating extensions of GR \cite{2009PhR...480....1A,2018PhRvD..98l4018N}.  Similar helicity-dependent propagation effects have also been parametrized for compact-binary tests \cite{2020PTEP.2020i3E01Y,2022PhRvD.106h4005W}.  We work in the circular-polarization basis,
\begin{align*}
 h_R(f)&=\frac{h_+(f)+i h_\times(f)}{\sqrt{2}},&
 h_L(f)&=\frac{h_+(f)-i h_\times(f)}{\sqrt{2}} .
\end{align*}
Phase birefringence is modeled by opposite phase shifts of the two helicities,
\begin{equation*}
 h_R(f)\rightarrow h_R(f)e^{+i\Delta(f,z)},\qquad
 h_L(f)\rightarrow h_L(f)e^{-i\Delta(f,z)} .
\end{equation*}
In the linear basis this is a rotation of the tensor-polarization basis,
\begin{align*}
 h_+^{\br}(f)&=h_+(f)\cos\Delta(f,z)-h_\times(f)\sin\Delta(f,z),\\
 h_\times^{\br}(f)&=h_+(f)\sin\Delta(f,z)+h_\times(f)\cos\Delta(f,z).
\end{align*}
The detector strain is
\begin{equation*}
 h_I(f)=F_I^+ h_+^{\br}(f)+F_I^\times h_\times^{\br}(f),
\end{equation*}
where \(I\) labels the detector and the antenna factors depend on sky position, polarization angle, and coalescence time.  The rotation is applied before detector projection.  It is therefore not equivalent to multiplying the detector-frame strain by a common phase.  This distinction is important for GW231123 because the event is observed by only the two LIGO detectors, so polarization, inclination, and distance are strongly correlated.

To connect with previous phenomenological searches, we start from the distance-dependent phase-rotation model
\begin{equation}
 \Delta(f,z)=\frac{1}{2}\beta_{\br}
 \left(\frac{f}{100\,\mathrm{Hz}}\right)^2
 \left[\frac{\Dbr(z)}{1000\,\mathrm{Mpc}}\right],
 \label{eq:delta_model}
\end{equation}
where \(\beta_{\br}\) is dimensionless.  The quadratic frequency dependence is a phenomenological choice rather than a unique prediction of one Lagrangian.  It captures the behavior that higher-frequency modes can accumulate a larger helicity-dependent phase while the overall effect grows with propagation distance.  Such parametrized frequency-dependent propagation effects are standard in modified-propagation tests \cite{2020PTEP.2020i3E01Y,2022PhRvD.106h4005W,2012PhRvD..85b4041M}.  We use
\begin{equation*}
 \Dbr(z)=\frac{c}{H_0}\int_0^z
 \frac{(1+z')\,\dd z'}{\sqrt{\Omega_M(1+z')^3+\Omega_\Lambda}},
\end{equation*}
for a flat \(\Lambda\)CDM cosmology; the factor \((1+z')\) accounts for the redshifting of the observed GW frequency along the path.

Directly sampling \(\beta_{\br}\) is not ideal for this event because a nearly frequency-independent rotation is degenerate with the polarization angle \(\psi\).  We therefore sample the band-differential rotation
\begin{equation*}
 \db\equiv \Delta(448\,\mathrm{Hz})-\Delta(20\,\mathrm{Hz}),
\end{equation*}
with prior \(\db\in[-\pi,\pi]\).  The rotation inserted into the waveform is
\begin{equation*}
 \Delta(f)=\db\,
 \frac{(f/100\,\mathrm{Hz})^2-(20/100)^2}{(448/100)^2-(20/100)^2},
\end{equation*}
so that \(\Delta(20\,\mathrm{Hz})=0\) and \(\Delta(448\,\mathrm{Hz})=\db\).  This keeps the frequency-dependent part of the rotation in the analysis band and removes the constant part that is most degenerate with \(\psi\).

For comparison with Eq.~\eqref{eq:delta_model}, we define the derived coefficient
\begin{equation}
 \beta_{\br}^{\rm derived}
 =\frac{2\db}{[\Dbr(z)/1000\,\mathrm{Mpc}]
 \left[(448/100)^2-(20/100)^2\right]} .
 \label{eq:beta_derived}
\end{equation}
This quantity is not independently sampled.  It is computed from posterior samples of \(\db\) and the inferred distance or redshift.  We therefore use \(\beta_{\br}^{\rm derived}\) only as an equivalent coefficient for comparison with the distance-normalized parametrization.  We also quote \(|\Delta(100\,\mathrm{Hz})|\), \(|\Delta(200\,\mathrm{Hz})|\), and \(|\Delta(448\,\mathrm{Hz})|\).  The GR limit is \(\db=0\), equivalently \(\beta_{\br}^{\rm derived}=0\).

\section{Data, priors, and likelihood}
\label{sec:analysis}

\begin{table*}[t]
\renewcommand{\arraystretch}{1.5}
\caption{Fiducial prior distributions. The XPHM and XO4a analyses use the wide mass-ratio and spin ranges shown here; NRSur7dq4 runs are restricted to the surrogate validity domain, with \(q\ge 1/6\) and spin magnitudes \(a_{1,2}\le0.8\). Angles are in radians.}
\label{tab:priors}
\begin{ruledtabular}
\begin{tabular}{lll}
Parameter & Prior & Comment \\
\hline
\(\mathcal{M}^{\detf}\) & uniform in \([80,220]~\Msun\) & detector-frame chirp mass \\
\(q=m_2/m_1\) & uniform in \([0.1,1]\) & phenomenological models; NRSur restricted to \([1/6,1]\) \\
\(a_1,a_2\) & uniform in \([0,0.99]\) & NRSur restricted to \([0,0.8]\) \\
\(\cos\theta_1,\cos\theta_2\) & uniform in \([-1,1]\) & isotropic spin tilts \\
\(\phi_{12},\phi_{JL}\) & uniform in \([0,2\pi)\) & periodic spin azimuths \\
\(D_L\) & source-frame volumetric prior, \([0.1,8]~\mathrm{Gpc}\) & Planck cosmology \\
\(\cos\theta_{JN}\) & uniform in \([-1,1]\) & isotropic binary orientation \\
\(\alpha\) & uniform in \([0,2\pi)\) & right ascension \\
\(\sin\delta\) & uniform in \([-1,1]\) & isotropic declination \\
\(\psi\) & uniform in \([0,\pi)\) & polarization angle \\
\(\phi_0\) & uniform in \([0,2\pi)\) & sampled coalescence phase \\
\(t_c\) & uniform in \([t_0-0.1,t_0+0.1]~\mathrm{s}\) & geocentric coalescence time \\
\(\db\) & uniform in \([-\pi,\pi]\) & band-differential birefringent rotation \\
\end{tabular}
\end{ruledtabular}
\end{table*}
We use the public GW231123 data products released with the LVK analysis \cite{2025ApJ...993L..25A}.  The fiducial data set consists of the Hanford strain after BayesWave glitch subtraction and the original Livingston strain.  BayesWave is a Bayesian wavelet-based method for modelling transient gravitational-wave signals and instrumental glitches in non-Gaussian detector noise \cite{2015CQGra..32m5012C}.  The likelihood uses the same 8 s analysis segment and the same frequency band, \(20\)--\(448\,\mathrm{Hz}\).  For each waveform family we use the corresponding public PSD estimate.  The strain data are resampled onto a common \(1024\,\mathrm{Hz}\) grid; this step fixes the numerical time grid and does not introduce an additional physical filtering choice.

For a hypothesis \(H\), the posterior is
\begin{equation*}
 p(\bm{\theta}|d,H)=\frac{\mathcal{L}(d|\bm{\theta},H)\pi(\bm{\theta}|H)}{Z_H}.
\end{equation*}
Here \(\mathcal{L}\) is the likelihood, \(\pi\) is the prior, and
\(Z_H=\int \mathcal{L}(d|\bm{\theta},H)\pi(\bm{\theta}|H)\,\dd\bm{\theta}\)
is the Bayesian evidence.  We use the standard stationary Gaussian likelihood,
\begin{equation*}
 \ln\mathcal{L}=-\frac{1}{2}\sum_I
 \left\langle d_I-h_I(\bm{\theta})\middle|d_I-h_I(\bm{\theta})\right\rangle
 +\mathrm{const.},
\end{equation*}
where \(\langle a|b\rangle\) denotes the usual PSD-weighted inner product over the analyzed detector frequency range.  The Bayes factor comparing the birefringent model with GR is
\begin{equation}
 \ln\mathcal{B}_{\br/\GR}=\ln Z_{\br}-\ln Z_{\GR}.
 \label{eq:bayes_factor}
\end{equation}
The symbols \(Z_{\GR}\) and \(Z_{\br}\) denote the evidences for the GR and birefringent hypotheses, respectively.  Absolute evidence values depend on common likelihood normalizations; only evidence differences computed with matched data conditioning, PSDs, priors, waveform family, and sampler settings are interpreted.  We use \textsc{bilby} and \textsc{dynesty} for parameter estimation and evidence calculation \cite{2019ApJS..241...27A,2020MNRAS.493.3132S}.  Distance marginalization is not used because the birefringent phase depends explicitly on distance or redshift.  The coalescence phase is sampled.

The priors are chosen to cover the public single-waveform posteriors while keeping the analysis close to the LVK source-property setup.  Detector-frame chirp mass and mass ratio are sampled directly.  Detector-frame component masses are derived from
\begin{align*}
 m_1^{\detf}&=\mathcal{M}^{\detf}\frac{(1+q)^{1/5}}{q^{3/5}},&
 m_2^{\detf}&=\mathcal{M}^{\detf}q^{2/5}(1+q)^{1/5}.
\end{align*}
Source-frame masses are then obtained from \(m_i^{\src}=m_i^{\detf}/(1+z)\).  Table~\ref{tab:priors} summarizes the priors.  The \textsc{NRSur7dq4} analysis is restricted to the surrogate validity domain, while XPHM and XO4a use the wider phenomenological-model ranges.

Our production runs use XPHM, XO4a, and NRSur7dq4.  XPHM is a frequency-domain phenomenological inspiral-merger-ringdown model with precession and higher modes \cite{2021PhRvD.103j4056P}.  XO4a is a newer phenomenological precessing higher-mode model \cite{2024PhRvD.109f3012T}.  NRSur7dq4 is a numerical-relativity surrogate for precessing binary black holes \cite{2019PhRvR...1c3015V}.  Because the surrogate has a finite calibration domain, the NRSur7dq4 result should be read as a robustness check within that domain, not as an identical prior-volume comparison with the two phenomenological models.

\section{Diagnostics and interpretation strategy}
\label{sec:diagnostics}

\begin{table*}[t]
\renewcommand{\arraystretch}{1.5}
\caption{Birefringent posterior summaries for the three waveform families. Values are medians with central 90\% credible intervals. Masses are in solar masses.}
\label{tab:source_results}
\begin{ruledtabular}
\begin{tabular}{lcccccc}
Waveform & \(m_1^{\src}\) & \(m_2^{\src}\) & \(M_{\rm tot}^{\detf}\) & \(q\) & \(D_L\,[\mathrm{Gpc}]\) & \(z\) \\
\hline
XPHM & $149.0^{+12.00}_{-11.74}$ & $94.23^{+15.86}_{-19.97}$ & $287.1^{+15.92}_{-20.53}$ & $0.632^{+0.110}_{-0.128}$ & $0.896^{+0.391}_{-0.299}$ & $0.179^{+0.068}_{-0.056}$ \\
XO4a & $140.0^{+21.78}_{-11.73}$ & $54.57^{+9.35}_{-8.13}$ & $315.7^{+14.16}_{-13.22}$ & $0.391^{+0.043}_{-0.071}$ & $3.81^{+1.17}_{-1.39}$ & $0.623^{+0.153}_{-0.197}$ \\
NRSur7dq4 & $133.1^{+12.70}_{-10.98}$ & $112.1^{+11.79}_{-13.75}$ & $305.8^{+16.90}_{-20.98}$ & $0.846^{+0.127}_{-0.151}$ & $1.24^{+0.709}_{-0.470}$ & $0.239^{+0.116}_{-0.083}$ \\
\end{tabular}
\end{ruledtabular}
\end{table*}

The analysis is not judged by a Bayes factor alone.  GW231123 is a waveform-systematics stress test: an extra parameter can improve the likelihood either because the data contain a physical propagation effect or because the parameter absorbs waveform mismatch.  We therefore apply three diagnostics.

First, the GR baseline for each waveform family must reproduce the corresponding public single-waveform posterior within sampling uncertainty.  If a GR run already misses the public mass--distance mode, a beyond-GR extension cannot be interpreted safely.

Second, detector-frame and source-frame quantities must be separated.  Source-frame masses depend on the inferred redshift.  A model can move \(m_1^{\src}\) and \(m_2^{\src}\) by changing \(D_L\), even if the detector-frame waveform morphology is not improved.  We therefore compare \(M_{\rm tot}^{\detf}\), \(q\), \(m_1^{\src}\), \(m_2^{\src}\), \(D_L\), \(\theta_{JN}\), and \(\psi\).

Third, a propagation interpretation requires stability.  A real birefringence signal should not appear only for one waveform family.  It should also survive reasonable frequency-cut, PSD, and prior checks, and it should be recoverable in injections.  In this paper, a \emph{constraint} means that \(\db=0\) and \(\beta_{\br}^{\rm derived}=0\) are compatible with the posterior and that the evidence is not waveform-robustly positive.  A \emph{birefringence-like systematic} means that a waveform family prefers the extra rotation parameter but the preference is not stable.  A \emph{candidate propagation anomaly} would require nonzero support that is stable across waveform families and validation tests.  Our results fall in the second category for XO4a and in the first category overall.

\begin{figure*}[t]
\centering
\includegraphics[width=\textwidth]{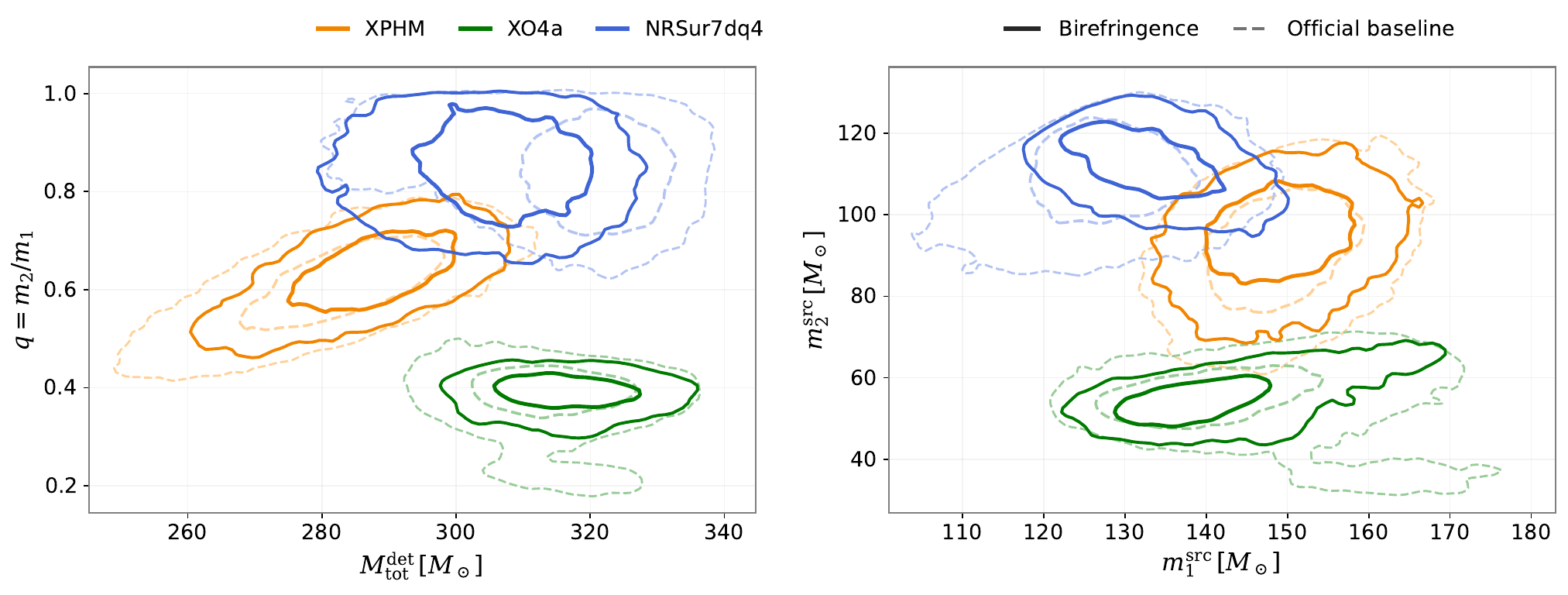}
\caption{Detector-frame and source-frame mass posteriors for the birefringent runs compared with the corresponding public single-waveform baselines. Solid contours show the birefringent analysis and dashed contours show the official baseline. The left panel uses \(M_{\rm tot}^{\detf}\) and \(q\); the right panel uses \(m_1^{\src}\) and \(m_2^{\src}\).}
\label{fig:mass_compare}
\end{figure*}

\section{Results}
\label{sec:results}

Table~\ref{tab:source_results} summarizes the birefringent source-parameter posteriors for the three waveform families. The waveform dependence remains substantial. XPHM and NRSur7dq4 both give source-frame total masses near \(240\)--\(250\Msun\), but their mass ratios are very different: XPHM favors \(q=0.632^{+0.110}_{-0.128}\), whereas NRSur7dq4 favors a near-equal-mass solution, \(q=0.846^{+0.127}_{-0.151}\). XO4a gives a more asymmetric and more distant solution, with \(q=0.391^{+0.043}_{-0.071}\) and \(D_L=3.81^{+1.17}_{-1.39}\,\mathrm{Gpc}\). Thus the birefringent degree of freedom does not collapse the three waveform families onto a common detector-frame morphology. This is visible in Fig.~\ref{fig:mass_compare}: the detector-frame \(M_{\rm tot}^{\detf}\)--\(q\) plane and the source-frame \(m_1^{\src}\)--\(m_2^{\src}\) plane both retain clear waveform-dependent structure.

\begin{table}[t]
\renewcommand{\arraystretch}{1.5}
\caption{Spin posterior summaries for the birefringent runs. Values are medians with central 90\% credible intervals. The NRSur7dq4 spin magnitudes are restricted by the surrogate validity domain, \(a_{1,2}\le 0.8\).}
\label{tab:spin_results}
\begin{ruledtabular}
\begin{tabular}{lcccc}
Waveform & \(a_1\) & \(a_2\) & \(\chi_{\rm eff}\) & \(\chi_p\) \\
\hline
XPHM &
$0.746^{+0.201}_{-0.252}$ &
$0.769^{+0.201}_{-0.535}$ &
$0.101^{+0.169}_{-0.261}$ &
$0.714^{+0.204}_{-0.223}$ \\
XO4a &
$0.921^{+0.059}_{-0.073}$ &
$0.449^{+0.465}_{-0.396}$ &
$0.330^{+0.175}_{-0.129}$ &
$0.814^{+0.081}_{-0.101}$ \\
NRSur7dq4 &
$0.719^{+0.074}_{-0.195}$ &
$0.722^{+0.072}_{-0.188}$ &
$0.181^{+0.233}_{-0.292}$ &
$0.635^{+0.127}_{-0.155}$ \\
\end{tabular}
\end{ruledtabular}
\end{table}

The spin posteriors strengthen the same conclusion. Table~\ref{tab:spin_results} shows that the three birefringent runs occupy different spin regions. XPHM allows broad high-spin support, with \(a_1=0.746^{+0.201}_{-0.252}\) and \(a_2=0.769^{+0.201}_{-0.535}\). XO4a prefers a very large primary spin, \(a_1=0.921^{+0.059}_{-0.073}\), together with a positive effective spin, \(\chi_{\rm eff}=0.330^{+0.175}_{-0.129}\), and strong precession, \(\chi_p=0.814^{+0.081}_{-0.101}\). NRSur7dq4 gives high but domain-limited spin magnitudes, \(a_1=0.719^{+0.074}_{-0.195}\) and \(a_2=0.722^{+0.072}_{-0.188}\). Thus the XO4a solution is not just the XPHM or NRSur7dq4 binary plus a propagation phase. It is a different waveform-dependent posterior mode involving spin, mass ratio, distance, inclination, and polarization.

The birefringence posterior constraints are given in Table~\ref{tab:beta}. The directly sampled parameter \(\db\) is broad for all three waveform families, with 90\% absolute limits near \(|\db|_{90}\simeq 2.8\,\mathrm{rad}\). This is close to the \([-\pi,\pi]\) prior scale, indicating that GW231123 does not tightly constrain the accumulated band-differential rotation itself. After mapping to the distance-normalized coefficient in Eq.~\eqref{eq:beta_derived}, the derived posteriors are centered near the GR value in all cases. The tightest derived limit comes from XO4a, \(|\beta_{\br}^{\rm derived}|_{90}=0.097\), because the XO4a posterior is driven to larger distances. XPHM and NRSur7dq4 give weaker limits, \(|\beta_{\br}^{\rm derived}|_{90}=0.378\) and \(0.273\), respectively. The corresponding limits on the effective rotation are similar across the waveform families: \(|\Delta(100\,\mathrm{Hz})|_{90}\simeq0.13\)--\(0.14\,\mathrm{rad}\), \(|\Delta(200\,\mathrm{Hz})|_{90}\simeq0.55\)--\(0.56\,\mathrm{rad}\), and \(|\Delta(448\,\mathrm{Hz})|_{90}\simeq2.8\,\mathrm{rad}\).

\begin{table*}[t]
\renewcommand{\arraystretch}{1.5}
\caption{Birefringence posterior constraints. The sampled parameter is \(\db=\Delta(448\,\mathrm{Hz})-\Delta(20\,\mathrm{Hz})\) with prior \([-\pi,\pi]\). The columns for \(\db\) and \(\beta_{\br}^{\rm derived}\) give medians with central 90\% credible intervals. The remaining columns give 90\% upper limits on absolute rotation quantities. All rotation angles are in radians.}
\label{tab:beta}
\begin{ruledtabular}
\begin{tabular}{lccccccc}
Waveform & \(\db\) prior & \(\db\) median and 90\% CI & \(\beta_{\br}^{\rm derived}\) median and 90\% CI & \(|\beta_{\br}^{\rm derived}|_{90}\) & \(|\Delta_{100}|_{90}\) & \(|\Delta_{200}|_{90}\) & \(|\Delta_{448}|_{90}\) \\
\hline
XPHM & $[-\pi,\pi]$ & $-1.23^{+3.51}_{-1.72}$ & $-0.147^{+0.388}_{-0.286}$ & $0.378$ & $0.136$ & $0.559$ & $2.83$ \\
XO4a & $[-\pi,\pi]$ & $0.124^{+2.70}_{-2.93}$ & $0.004^{+0.092}_{-0.102}$ & $0.097$ & $0.135$ & $0.558$ & $2.82$ \\
NRSur7dq4 & $[-\pi,\pi]$ & $-0.588^{+3.20}_{-2.29}$ & $-0.051^{+0.273}_{-0.254}$ & $0.273$ & $0.134$ & $0.551$ & $2.79$ \\
\end{tabular}
\end{ruledtabular}
\end{table*}

\begin{figure*}[t]
\centering
\includegraphics[width=\textwidth]{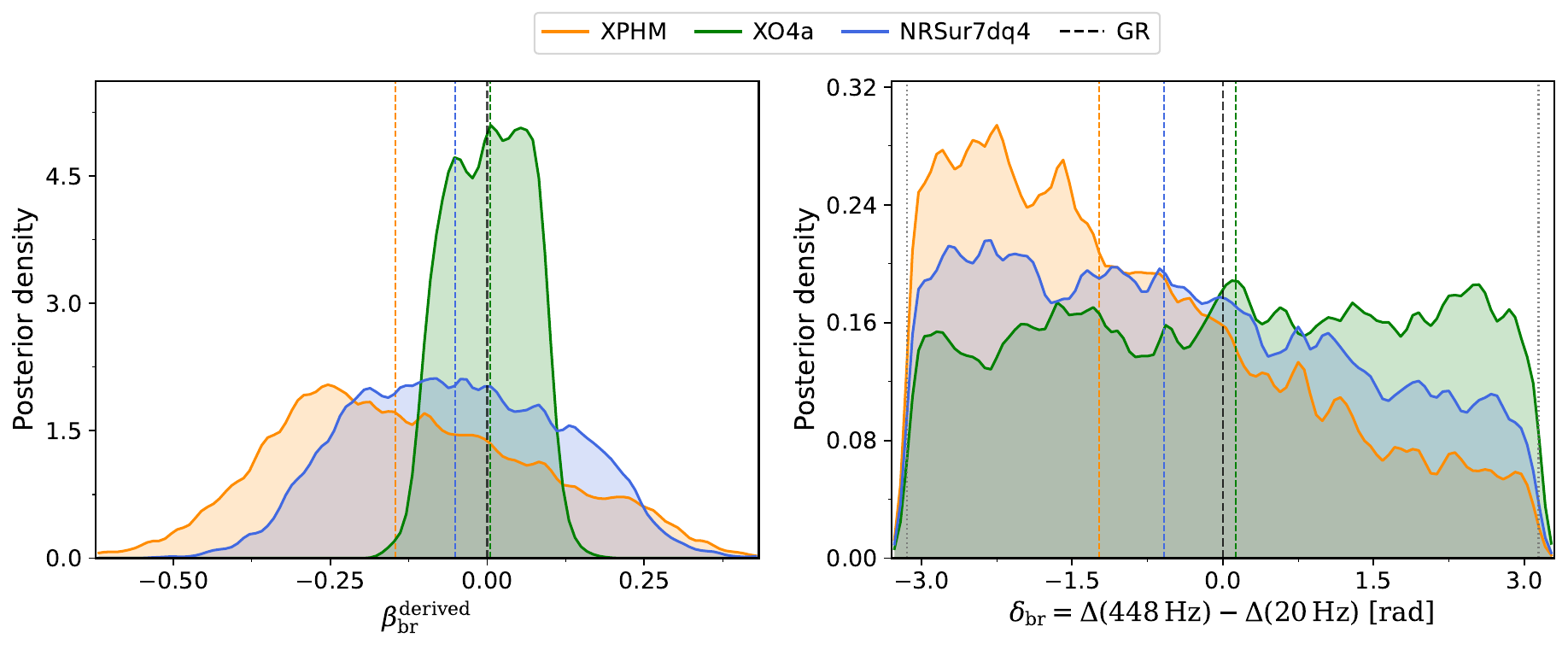}
\caption{Marginalized posteriors for the derived propagation coefficient \(\beta_{\br}^{\rm derived}\) and for the directly sampled band-differential rotation \(\db\). The vertical dashed line marks the GR value. The broad \(\db\) posteriors show that the rotation across the full analysis band is weakly constrained; the derived \(\beta_{\br}^{\rm derived}\) distributions should therefore be interpreted as equivalent-coefficient summaries rather than direct detections.}
\label{fig:beta_delta}
\end{figure*}

\begin{figure*}[t]
\centering
\includegraphics[width=1\textwidth]{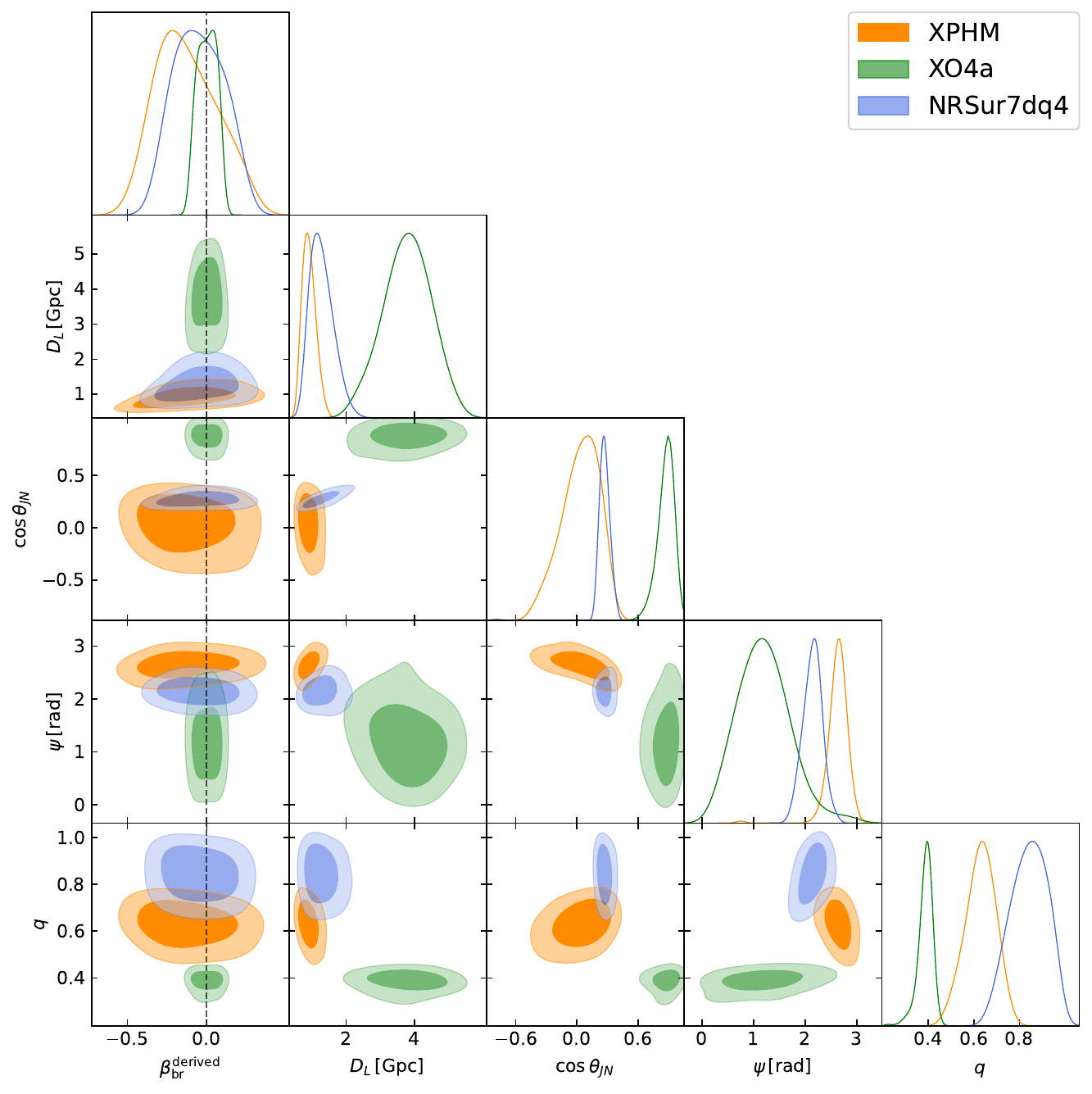}
\caption{Posterior correlations of \(\beta_{\br}^{\rm derived}\) with luminosity distance, inclination, polarization angle, and mass ratio. The vertical dashed line denotes the GR value. The figure illustrates that the apparent sharpness of \(\beta_{\br}^{\rm derived}\), especially for XO4a, is tied to distance--orientation structure in the posterior.}
\label{fig:corner}
\end{figure*}

Figure~\ref{fig:corner} shows the main degeneracies. The derived coefficient is correlated with the luminosity distance because Eq.~\eqref{eq:beta_derived} divides the sampled rotation by the propagation kernel. The XO4a solution lies at much larger distance and near face-on orientation, producing the narrowest \(\beta_{\br}^{\rm derived}\) posterior but not a correspondingly well-constrained \(\db\). The XPHM and NRSur7dq4 runs occupy different regions of the \((D_L,\cos\theta_{JN},\psi,q)\) space. These correlations show that the propagation parameter is entangled with the extrinsic geometry and the mass-ratio mode rather than acting as an isolated phase correction.

\begin{figure}[t]
\centering
\includegraphics[width=\columnwidth]{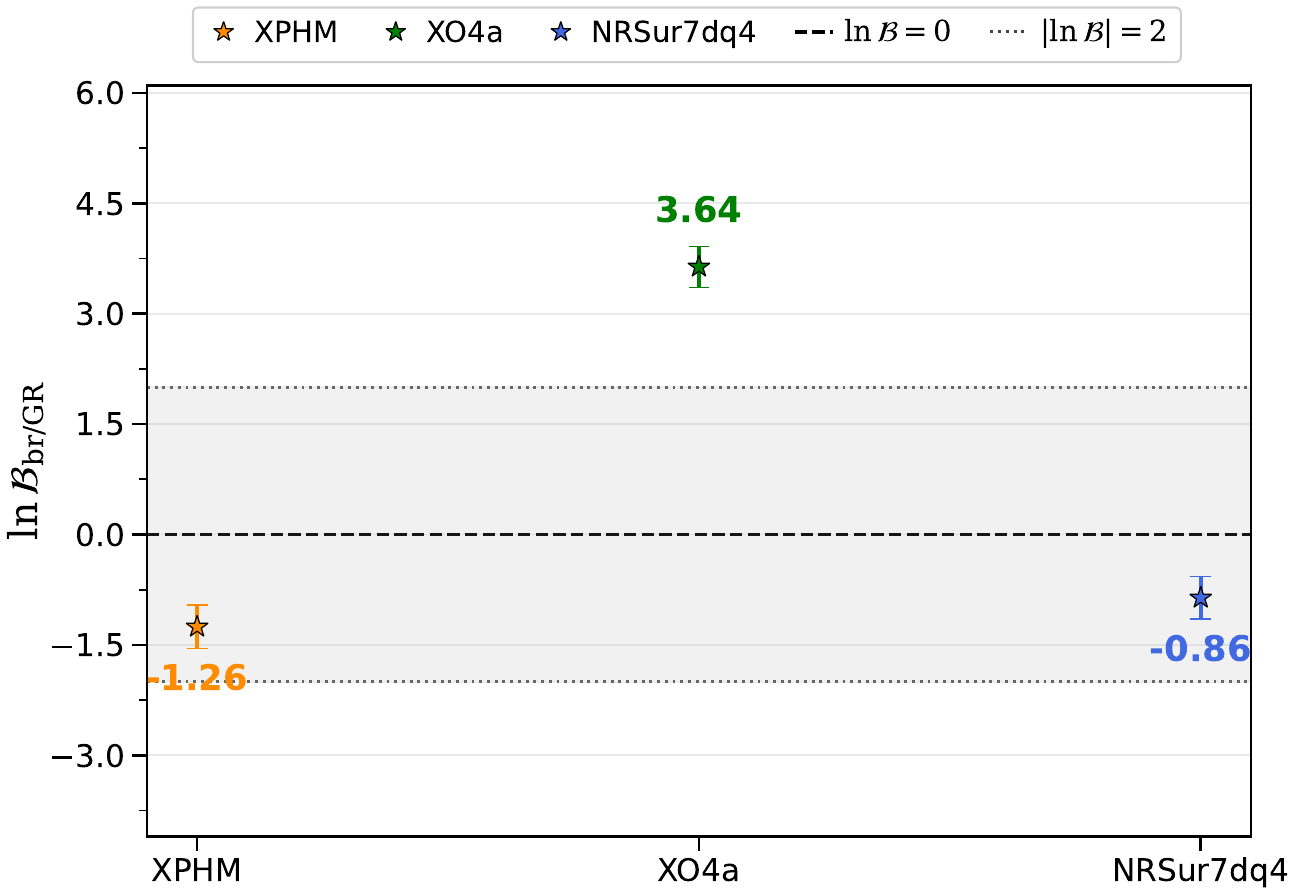}
\caption{Bayes factors for the birefringent model relative to the corresponding GR baseline. The shaded interval \(|\ln\mathcal{B}|<2\) marks weak evidence. Only XO4a shows positive support for the extended model; the absence of comparable support in XPHM and NRSur7dq4 indicates that the effect is waveform dependent.}
\label{fig:evidence}
\end{figure}

Fig.~\ref{fig:evidence} show the model comparison. XPHM gives \(\ln\mathcal{B}_{\br/\GR}=-1.26\pm0.30\), and NRSur7dq4 gives \(-0.86\pm0.29\); both values mildly disfavor the birefringent extension. XO4a gives \(+3.64\pm0.28\), so the extended model is favored within that waveform family. This is a real result of the XO4a analysis, but it is not reproduced by XPHM or NRSur7dq4. We therefore do not interpret it as waveform-robust evidence for parity-violating propagation. The direct interpretation is that the polarization-rotation parameter is sensitive to the waveform-dependent spin--mass-ratio--distance--orientation mode selected by XO4a.

Taken together, the posterior and evidence results support a clear conclusion. GW231123 does not provide a waveform-robust detection of gravitational-wave birefringence. The data allow a broad band-differential rotation, but the directly sampled \(\db\) remains weakly constrained. The derived \(\beta_{\br}^{\rm derived}\) is consistent with zero for all three waveform families. The only positive evidence occurs for XO4a, and it coincides with a very different distance--inclination--mass-ratio mode. This is precisely the situation in which a beyond-GR propagation parameter can act as a diagnostic of waveform systematics rather than as a physical discovery.

This result is different from a catalog-level birefringence constraint. Catalog analyses such as Refs.~\cite{2020PTEP.2020i3E01Y,2022PhRvD.106h4005W} combine many events to constrain a common propagation parameter. They are powerful for population bounds, but they are not designed to diagnose why a single short, massive event gives different answers under different waveform families. The amplitude-birefringence analysis of Ref.~\cite{2022PhRvD.106d4067O} also probes a different effect: helicity-dependent amplification rather than the phase rotation used here. The novelty of our analysis is therefore the event-level comparison. GW231123 shows explicitly that a birefringence-like parameter can be favored by one waveform family and disfavored by others.

\section{Discussion}
\label{sec:discussion}

The result is straightforward.  GW231123 does not give a waveform-robust detection of polarization birefringence.  It does show that a birefringence parameter can respond strongly to waveform systematics in a short, high-mass event.  This diagnostic point, rather than the numerical upper limit alone, is the main output of the paper.

First, the event gives an upper limit on an equivalent distance-normalized coefficient, but it does not tightly measure the band-differential rotation itself.  The sampled parameter \(\db\) remains broad, with \(|\db|_{90}\simeq2.8\,\mathrm{rad}\), close to the \([-\pi,\pi]\) prior scale.  The derived parameter \(\beta_{\br}^{\rm derived}\) is narrower because it divides the same sampled rotation by the propagation-distance kernel and the frequency lever arm.  Therefore a compact \(\beta_{\br}^{\rm derived}\) posterior is not a direct measurement of accumulated birefringent rotation.  It is an equivalent-coefficient summary of a weakly constrained rotation angle.

Second, the source posterior remains waveform dependent after adding the birefringence parameter.  XO4a prefers a more distant, more asymmetric, high-primary-spin solution than XPHM or NRSur7dq4.  Because source-frame masses depend on redshift, this distance shift changes the astrophysical interpretation.  It does not by itself prove that the detector-frame waveform is better reconstructed.  This is why the detector-frame \(M_{\rm tot}^{\detf}\)--\(q\) panel in Figure~\ref{fig:mass_compare} is central to the paper.

Third, the evidence result is mixed, not positive overall.  XO4a alone gives \(\ln\mathcal{B}_{\br/\GR}=3.64\pm0.28\), while XPHM and NRSur7dq4 give negative Bayes factors.  The error bars are nested-sampling evidence errors; the larger uncertainty is the waveform dependence itself.  A physical propagation effect should not be supported by only one waveform family while being disfavored by the others.  The XO4a result is therefore best interpreted as a waveform-dependent birefringence-like response.  It points to a degeneracy involving spin, mass ratio, distance, inclination, and polarization, not to a universal helicity-dependent propagation effect.  The NRSur7dq4 result should be read with the additional caveat that the surrogate is restricted to \(a_{1,2}\le0.8\); it does not directly exclude the high-primary-spin mode preferred by XO4a, but it does show that the same birefringence preference is not recovered in the surrogate-calibrated region.

This interpretation is consistent with earlier work.  Catalog analyses of gravitational-wave birefringence combine many events to constrain a common propagation parameter \cite{2020PTEP.2020i3E01Y,2022PhRvD.106h4005W}.  Amplitude-birefringence studies in Chern--Simons-like theories constrain helicity-dependent amplification rather than the phase rotation considered here \cite{2022PhRvD.106d4067O}.  The LVK GWTC-3 tests of GR find no population-level evidence for modified dispersion, non-GR polarizations, or other deviations from GR \cite{2025PhRvD.112h4080A}.  Our analysis asks a different question: for one difficult high-mass event, does a propagation parameter stay stable when the waveform family changes?  For GW231123 the answer is no.

This is why GW231123 is complementary to clean, high-SNR events such as GW250114 \cite{2025PhRvL.135k1403A}.  A loud, nearly equal-mass, low-spin event is the natural choice for a sharp null test and for black-hole spectroscopy.  GW231123 is less clean and less constraining as a standalone upper-limit event.  Its advantage is that it exposes a failure mode: a beyond-GR propagation parameter can look useful in one waveform family and disappear in another.  Future propagation searches should treat such behavior as a systematics warning, not as a discovery.

The comparison with lensing and overlapping-signal interpretations is also important.  Recent work has argued that lensing-like distortions, microlensing, and overlapping signals can produce frequency-dependent structure in GW231123-like short signals \cite{2025arXiv251216916C,2025arXiv251217550H,2025arXiv251217631G,2025arXiv250908298L,2025arXiv251219077C,2025arXiv251219118S}.  These effects are not equivalent to birefringence because they do not generically rotate the tensor-polarization basis.  But with only two detectors, different physical distortions can project onto similar likelihood improvements.  A birefringence claim for GW231123 would therefore need to survive explicit comparison with lensing-like and overlap-like alternatives.

Several checks are still needed before making a stronger claim.  The analysis should be repeated with TPHM and SEOBNRv5PHM to match the full LVK waveform set.  Single-detector and frequency-cut runs should test whether the XO4a preference is driven by a particular detector or frequency interval.  GR injections with high spins and precession should verify that waveform mismatch does not generate a false \(\db\) signal.  Finally, birefringent injections should demonstrate recovery of the sampled \(\db\) parameter when the assumed polarization rotation is really present.

\section{Conclusion}
\label{sec:conclusion}

We tested gravitational-wave phase birefringence with GW231123 using XPHM, XO4a, and NRSur7dq4.  We sampled the band-differential rotation \(\db\in[-\pi,\pi]\) and reported the derived distance-normalized coefficient \(\beta_{\br}^{\rm derived}\).  The posterior constraints are consistent with GR: \(|\beta_{\br}^{\rm derived}|_{90}=0.378\), \(0.097\), and \(0.273\) for XPHM, XO4a, and NRSur7dq4, respectively.  The directly sampled rotation remains broad, with \(|\db|_{90}\simeq2.8\,\mathrm{rad}\). Thus GW231123 does not tightly measure the accumulated birefringent rotation across the detector band.

The evidence result is the central finding.  XPHM and NRSur7dq4 do not favor the birefringent extension, while XO4a does.  The positive XO4a Bayes factor occurs together with a different mass-ratio, distance, orientation, and spin solution.  Because the same preference is absent in the other waveform families, it is not evidence for a waveform-robust parity-violating propagation effect.

The main output of the paper is therefore not a detection claim and not the strongest possible upper limit.  The main output is a diagnostic result: GW231123 shows explicitly that a polarization-rotation parameter can behave as a waveform-dependent nuisance direction in a short, high-mass, high-spin event.  A propagation parameter that is significant in only one waveform family should be treated as a systematics diagnostic until it passes waveform, frequency-cut, detector, and injection tests.

\section*{Acknowledgments}
We thank the LIGO Scientific, Virgo, and KAGRA Collaborations for making GW231123 data products publicly available. This work uses \textsc{bilby}, \textsc{gwpy}, \textsc{lalsuite}, \textsc{dynesty}, and public posterior samples released with the LVK GW231123 analysis.
This work was supported by National Key R$\&$D Program of China (No. 2024YFC2207400), and (Grant No. 2022YFC2204602).

\bibliographystyle{apsrev4-2}

\bibliography{ref}
\end{document}